\documentclass[reprint,superscriptaddress,showpacs,amsmath,amssymb,aps,pra]{revtex4-2}
\usepackage{pifont}
\usepackage{times}
\usepackage{amssymb}
\usepackage{graphicx}
\usepackage{dcolumn}
\usepackage{dsfont}
\usepackage{bm}
\usepackage{subfigure}
\usepackage[ruled]{algorithm2e}
\usepackage{hyperref}
\usepackage{appendix}
\usepackage{lipsum}

\hypersetup{colorlinks=true, citecolor=blue, urlcolor=blue, linkcolor=blue}

\newtheorem{theorem}{Theorem}

\newtheorem{corollary}{Corollary}
\newtheorem{definition}{Definition}

\makeatletter

\newcommand{\Rmnum}[1]{\expandafter\@slowromancap\romannumeral #1@}

\input amssym.def

\begin{document}
	\title{Imaginarity as a Resource within Quantum Coherence: Geometric Decomposition and Operational Conversion}
	
	\author{Meng-Li Guo}
	\affiliation{School of Science, East China University of Technology, Nanchang 330006, China}
	\author{Nannan Guan}
	\affiliation{School of Science, East China University of Technology, Nanchang 330006, China}
	\author{Bo Li}
	\email{libobeijing2008@163.com}
	\affiliation{School of Computer and Computing Science, Hangzhou City University, Hangzhou 310015, China}
	\author{Bin Hu}
	\email{bhu@ecut.edu.cn}
	\affiliation{School of Science, East China University of Technology, Nanchang 330006, China}
	\author{Shao-Ming Fei}
	\email{feishm@cnu.edu.cn}
	\affiliation{School of Mathematical Sciences, Capital Normal University, Beijing 100048, China}

	\begin{abstract}
		We establish a rigorous framework that identifies imaginarity as a fundamental resource inherent in quantum coherence. By means of a geometric decomposition, we partition coherence into distinct imaginarity and residual components, thereby revealing a universal hierarchical relationship among these resources.
		For bipartite systems, this decomposition provides explicit bounds on the extent to which nonlocal correlations and imaginarity limit local coherence generation. Furthermore, we devise an explicit operational protocol that converts imaginarity into usable coherence, demonstrating the direct interconvertibility of these resources under physically admissible operations.
		The dynamical evolution under diagonal Hamiltonians is fully characterized, showing that while total coherence is conserved, imaginarity and residual coherence exhibit complementary oscillations.
		Our results provide a rigorous geometric and operational characterization of imaginarity as a fundamental constituent of quantum coherence, offering concrete insights for resource management in distributed quantum technologies.
		The geometric framework and theoretical bounds established herein are fully general, while the explicit conversion protocol and dynamical analysis serve as a compelling proof-of-principle demonstration in qubit systems.
	\end{abstract}
	
	\maketitle
	
	\section{Introduction}
	Quantum resource theories offer a rigorous framework for quantifying the nonclassical features that enable advantages in information processing, metrology, and thermodynamics. Within this paradigm, quantum coherence, characterized by the existence of superpositions in a preferred basis, has been established as a fundamental resource. Its critical role is demonstrated across diverse domains, including enhanced thermodynamic processes at low temperatures~\cite{y1,y2,Lostaglio2015}, efficient energy transport in quantum biological systems~\cite{y6,y8,y11}, novel quantum phenomena in nanoscale physics~\cite{y13}, and the generation of nonclassical states in quantum optics~\cite{y14,y15,a2}. Formal quantification within resource-theoretic frameworks~\cite{Baumgratz,Girolami,Pires,Streltsov} has revealed profound connections to quantum speed limits~\cite{Marvian2016,Meng2023}, thermodynamic advantages surpassing free-energy constraints~\cite{Lostaglio2015}, and the activation of nonclassical correlations~\cite{Chitambar2016,Piani2011}. This solidifies coherence as an essential ingredient for protocols ranging from secure communication and precision metrology to advanced computational tasks~\cite{spm,chs,jbdv,ad,easm,em,eg,irp,yxlc,ula,aer,mmtc,cmst,bukf}.
	
	Parallel to this, the resource character of imaginarity, rooted in the essential complex phase structure of quantum mechanics, has garnered significant theoretical and experimental interest. Foundational no-go theorems established the insufficiency of real-number-only formulations~\cite{Wootters37,Hardy38,Aleksandrova39}, and this conclusion has been decisively confirmed by photonic network experiments that violated the real-valued correlation bound with a significance exceeding five standard deviations~\cite{Renou2,Chen3,Li4}. The formal resource theory of imaginarity~\cite{Hickey1,Wu1,Wu17,Gao2023,Guo2024} quantifies this feature through rigorous monotones, and its operational relevance extends well beyond formal quantification. In quantum channel discrimination, imaginarity enables unambiguous identification of certain channels without requiring ancillas under real operations~\cite{KangDa}. In multiparameter estimation, it is essential for attaining Heisenberg-limited precision, a regime intrinsically inaccessible to real quantum systems~\cite{Miyazaki}. In state discrimination tasks with cryptographic implications, the complex phase structure can induce strong nonlocality in sets of orthogonal states, thereby providing a security advantage tied directly to the presence of imaginarity~\cite{Wei}. The physical distinction between coherence and imaginarity is also experimentally tangible: coherence measures superposition with respect to a preferred basis, whereas imaginarity captures the complex phase content that survives when all real components are diagonalized. In linear optical implementations, generating complex amplitudes requires additional wave plates compared with real operations, providing a direct operational cost measure for imaginarity resources~\cite{Wu1}. Recent advances continue to refine imaginarity quantification through convex roof extensions~\cite{Xue} and its characterization for Gaussian states~\cite{XuG}.
	
	Despite their independent development, the structural relationship between coherence and imaginarity remains a deep and unresolved question. Pioneering works established the resource-theoretic framework for imaginarity and its operational characterizations~\cite{Hickey1}. Subsequent studies systematically advanced imaginarity quantification, state transformations, and distributed applications~\cite{Wu1,Wu17}, while recent work explored mixed-state quantification from a pure-state perspective~\cite{DuBai}. These efforts have firmly established imaginarity as an independent resource. However, a fundamental question persists: What is the precise structural and geometric relationship between coherence and imaginarity? In distributed quantum systems, can the imaginarity resource in one subsystem be operationally converted into coherence in another, particularly under restricted operations? What are the dynamical constraints governing their interconversion under physical evolutions, and what are the fundamental limits to this conversion efficiency?
	
	This gap motivates our central breakthrough: the introduction of the novel concept of \emph{residual coherence}, defined as the metric distance between the closest real and closest incoherent states. This concept reveals a universal geometric decomposition where total coherence is bounded by the sum of imaginarity and residual coherence. This inequality demonstrates that coherence can be partitioned into a pure imaginarity component and a ``residual" component associated with real-part structure and basis misalignment. This not only clarifies imaginarity as a proper sub-resource of coherence but, more importantly, provides an operationally interpretable conversion cost, as residual coherence quantifies the inherent overhead from reference-frame mismatch when converting imaginarity into coherence. This decomposition, not captured in prior literature, explains from geometric first principles why coherence cannot always be fully attributed to imaginarity, thereby addressing a critical gap in the resource hierarchy.
	
	To operationalize this insight, our advances manifest on three interconnected fronts, establishing a coherent theoretical architecture from static decomposition to operational conversion and dynamical constraints. First, we generalize the decomposition to bipartite systems, explicitly revealing how nonlocal correlations and remote imaginarity jointly constrain local coherence generation. Second, we devise an explicit operational protocol, complete with quantum circuit implementation, that actively converts imaginarity in one subsystem into usable coherence in another under real operations, proving an exact conversion equality. Third, we fully characterize the complementary dynamical evolution under diagonal Hamiltonians, uncovering a conservation law for total coherence alongside oscillations between imaginarity and residual coherence, while deriving general conversion limits that quantify the intrinsic efficiency of resource activation. Collectively, this work provides a comprehensive geometric and operational understanding of imaginarity as a constitutive element of quantum coherence, with immediate implications for resource management in distributed quantum technologies.

	\section{Geometric Framework for Coherence-Imaginarity Relations}
	Consider a $d$-dimensional Hilbert space $\mathcal{H}$ with fixed orthonormal basis $\{|j\rangle\}_{j=0}^{d-1}$. Let $\mathfrak{D}(\mathcal{H})$ denote the set of density operators on $\mathcal{H}$. Within the resource theory of imaginarity, the free states constitute the set of real density matrices:
	\begin{eqnarray*}
		\mathcal{F} = \{\rho \in \mathfrak{D}(\mathcal{H}) : \langle m|\rho|n \rangle \in \mathbb{R}, \, m,n = 0,1,\ldots,d-1\}.
	\end{eqnarray*}
	The free operations are quantum channels given by real Kraus operators $\{K_j\}$ satisfying $\sum_j K_j^\dagger K_j = \mathbb{I}$ and $\langle m|K_j|n \rangle \in \mathbb{R}$ for all $j,m,n$, where $\mathbb{I}$ denotes the identity operator.
	The quantum imaginarity $\mathcal{M}(\rho)$ of a state $\rho$ is defined by~\cite{Hickey1}
	\begin{eqnarray}\label{imaginaritydef}
		\mathcal{M}(\rho) = \min_{\sigma \in \mathcal{F}} \mathcal{D}(\rho, \sigma),
	\end{eqnarray}
	where $\mathcal{D}$ is a metric on the quantum state space.
	Within the resource theory of imaginarity, a valid measure $\mathcal{M}$ defined via a contractive distance $\mathcal{D}$ must be faithful, satisfying $\mathcal{M}(\rho) \geq 0$ with equality if and only if $\rho \in \mathcal{F}$, and monotonic under any real channel $\varepsilon$, meaning $\mathcal{M}(\varepsilon(\rho)) \leq \mathcal{M}(\rho)$. Moreover, strong monotonicity demands that for any real instrument $\{K_j\}$ with outcome probabilities $p_j = \mathrm{Tr}(K_j \rho K_j^\dagger)$ and post-measurement states $\rho_j = K_j \rho K_j^\dagger / p_j$, the expected imaginarity does not increase, i.e., $\sum_j p_j \mathcal{M}(\rho_j) \leq \mathcal{M}(\rho)$. Convexity further requires $\mathcal{M}(\sum_j p_j \rho_j) \leq \sum_j p_j \mathcal{M}(\rho_j)$ for any ensemble $\{p_j, \rho_j\}$.
	
	The quantum coherence $\mathcal{C}(\rho)$ of a state $\rho$ is defined as~\cite{Baumgratz}
	\begin{eqnarray}\label{coherencedef}
		\mathcal{C}(\rho) = \min_{\delta \in \mathcal{I}} \mathcal{D}(\rho, \delta),
	\end{eqnarray}
	where $\mathcal{I}$ denotes the set of incoherent states, the states that are diagonal with respect to the basis $\{|j\rangle\}_{j=0}^{d-1}$. Since $\mathcal{I} \subset \mathcal{F}$ (as diagonal density matrices are real), one has
	$\mathcal{M}(\rho) \leq \mathcal{C}(\rho)$.
	
	Specifically, while imaginarity quantifies the phase-dependent interference and coherence characterizes the superposition, their interplay imposes fundamental limitations on quantum protocols. Understanding their interplay is crucial for quantum resource theory. To this end, we introduce the concept of \emph{residual coherence}:
	
	\begin{definition}\label{definition 1}
		Let $\mathcal{I}$ and $\mathcal{F}$ be the sets of incoherent and real states, respectively. For any metric $\mathcal{D}$ satisfying the following properties:
		(i) \emph{Positivity}: $\mathcal{D}(\rho, \sigma) > 0$ whenever $\rho \neq \sigma$, and $\mathcal{D}(\rho, \rho) = 0$;
		(ii) \emph{Symmetry}: $\mathcal{D}(\rho, \sigma) = \mathcal{D}(\sigma, \rho)$;
		(iii) \emph{Triangle inequality}: $\mathcal{D}(\rho, \sigma) + \mathcal{D}(\sigma, \tau) \geq \mathcal{D}(\rho, \tau)$.
		We define the \emph{residual coherence} as
		\begin{eqnarray}
			\mathcal{C}_R(\rho) = \mathcal{D}(\rho_R, \rho_d),
		\end{eqnarray}
		where $\rho_R$ and $\rho_d$ represent the optimal solutions to Eqs.~(\ref{imaginaritydef}) and (\ref{coherencedef}), respectively. Both $\rho_d$ and $\rho_R$ are implicitly dependent on the input state $\rho$.  
		By the positivity property of the metric $\mathcal{D}$, it follows that $\mathcal{C}_R(\rho) \geq 0$, with equality if and only if $\rho_R = \rho_d$.
		If the optimal states are not unique, $\mathcal{C}_R(\rho)$ is defined as the infimum of $\mathcal{D}(\rho_R, \rho_d)$ over all admissible pairs drawn from the respective sets of minimizers.
	\end{definition}
	
\begin{figure}[t]
		\centering
		\includegraphics[width=0.8\columnwidth]{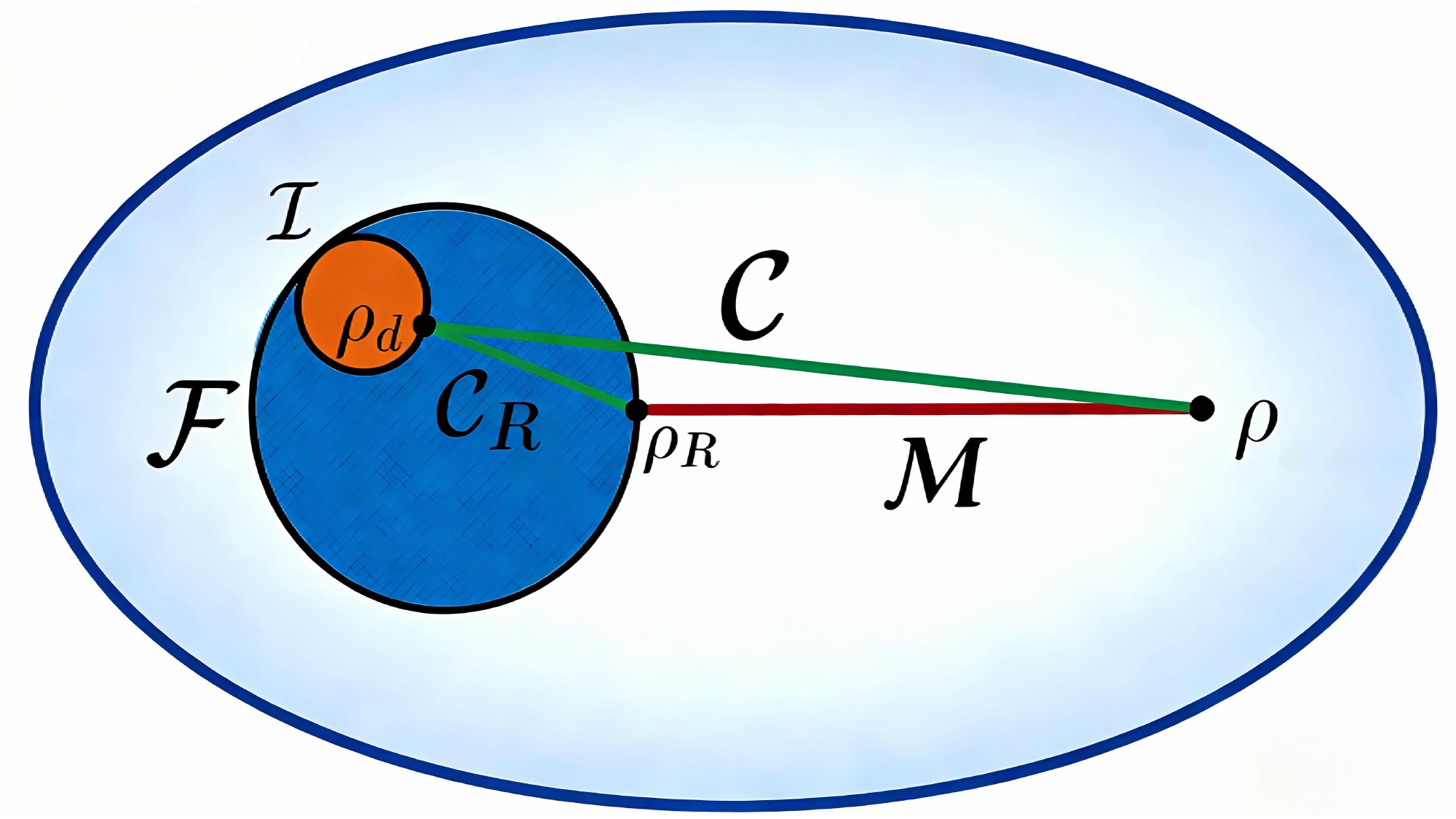}
		\caption{Geometric scheme of quantum resource measures. The total coherence of a state is bounded by the sum of its imaginarity (distance to nearest real state) and the residual coherence (distance between real and incoherent reference states), showing that imaginarity is a resource strictly contained within coherence.}
		\label{IM2}
	\end{figure}

	The triangle inequality satisfied by to $\mathcal{D}$ yields $\mathcal{C}(\rho)=\mathcal{D}(\rho, \rho_d)
	\leq \mathcal{D}(\rho,\rho_R)+\mathcal{D}(\rho_R,\rho_d)
	=\mathcal{M}(\rho)+\mathcal{C}_R(\rho)$. Therefore, we have the following fundamental trade-off relation among $\mathcal{C}$, $\mathcal{M}$ and $\mathcal{C}_R$, see Fig.~\ref{IM2}.
	
	\begin{theorem}\label{Theorem 1}
		For any quantum state $\rho$ and metric $\mathcal{D}$ satisfying properties (i)--(iii),
		\begin{eqnarray}\label{main}
			\mathcal{C}(\rho) \leq \mathcal{M}(\rho) + \mathcal{C}_R(\rho).
		\end{eqnarray}
	\end{theorem}
	
	We verify the validity of (\ref{main}) by using the trace distance, 
	which satisfies the properties (i)--(iii).
	Consider the qubit state,
	\begin{eqnarray*}
		\rho=\begin{pmatrix}
			\frac{1}{2} & x - iy \\
			x + iy & \frac{1}{2}
		\end{pmatrix}, \quad x,y \geq 0,~ x^2 + y^2 \leq \tfrac{1}{4}.
	\end{eqnarray*}
	Let the coherence be quantified by the modified trace norm $\mathcal{C}'_{\mathrm{tr}}(\rho) = \min_{\lambda \geq 0, \delta \in \mathcal{I}} \|\rho - \lambda \delta\|_{\mathrm{tr}}$, the imaginarity by $\mathcal{M}_{\mathrm{tr}}(\rho) = \min_{\sigma \in \mathcal{F}} \|\rho - \sigma\|_{\mathrm{tr}} = \frac{1}{2} \|\rho - \rho^T\|_{\mathrm{tr}}$, and $\mathcal{C}_{R_{\mathrm{tr}}}(\rho)=\|\rho_R-\rho_d\|_{\mathrm{tr}}$.
	Direct computation yields $\mathcal{C}'_{\mathrm{tr}}=\sqrt{x^2+y^2}$, $\mathcal{M}_{\mathrm{tr}}(\rho) =y$ and $\mathcal{C}_{R_{\mathrm{tr}}}(\rho) =x$, thereby confirming that the inequality $\mathcal{C}_{\mathrm{tr}}(\rho) \leq \mathcal{M}_{\mathrm{tr}}(\rho) + \mathcal{C}_{R_{\mathrm{tr}}}(\rho)$ holds.
	
	For mixed quantum state $\rho = \lambda |\psi\rangle\langle\psi| + \frac{1-\lambda}{d} \mathbb{I}_d$, 
	where $|\psi\rangle = \frac{1}{\sqrt{2}}(|1\rangle + i|2\rangle)$ and $0 \leq \lambda \leq 1$, we obtain
	$\mathcal{C}'_{\mathrm{tr}}(\rho)=\frac{\lambda}{2}$, $\mathcal{M}_{\mathrm{tr}}(\rho)=\frac{\lambda}{2}$ and $\mathcal{C}_{R_{tr}}(\rho)=0$.
	We have $\mathcal{C}_{tr}(\rho) = \mathcal{M}_{tr}(\rho)$. Since the quantum state $\rho$ is a pure imaginary state (i.e., the real parts of all off-diagonal elements are zero), we have $\rho_R = \rho_d$. Consequently, $\mathcal{C}_R(\rho) = 0$. the inequality in Theorem~1 reduces to an equality.
	
	\begin{corollary}\label{Corollary 1}
		For pure imaginary state $\rho$, it holds that $\mathcal{C}(\rho) = \mathcal{M}(\rho)$.
	\end{corollary}
	More generally, Eq.~(4) is saturated if and only if there exists an optimizing pair $(\rho_R,\rho_d)$ satisfying $D(\rho,\rho_d)=D(\rho,\rho_R)+D(\rho_R,\rho_d)$. In particular, every real state also saturates Eq.~(4), since $\rho_R=\rho$, $M(\rho)=0$, and $C_R(\rho)=C(\rho)$. For the qubit trace-norm family considered above, saturation is equivalent to $xy=0$, while the inequality is strict for $x>0$ and $y>0$.
	
	The inequality $\mathcal{C} \leq \mathcal{M} + \mathcal{C}_R$ reveals a fundamental geometric constraint on the quantum state space. From a physical perspective, $\mathcal{C}_R(\rho)$ arises from the basis mismatch between the closest real state $\rho_R$ and the closest incoherent state $\rho_d$. This geometric separation reflects an \textit{orthogonality penalty} when the optimal reference frames for phase-dependent (imaginarity) and basis-dependent (coherence) resources differ. Notably, this provides a clear explanation for Corollary \ref{Corollary 1}: the absence of real components eliminates any basis misalignment, resulting in $\mathcal{C}_R(\rho) = 0$. This mechanism is fundamentally distinct from the coherence factorization reported in Ref.~\cite{Zhao}, which relates coherence to metrological precision bounds, yet it is conceptually consistent with the framework for quantifying imaginarity developed in Ref.~\cite{DuBai}. Specifically, when $\mathcal{C}_R > 0$, the coexistence of these quantum resources enables access to quantum protocols that are not realizable within real quantum systems~\cite{Wu1}, while simultaneously introducing conversion limitations quantified by the distance $\mathcal{D}(\rho_R, \rho_d)$.
	
	On the other hand, in distributed quantum information processing and quantum network architectures, a fundamental challenge arises from the constraints on resource distribution across subsystems. Particularly under real operations, which preserve the set of real states, we face the critical question: To what extent can the imaginarity resource in one subsystem be harnessed to generate coherence in another spatially separated subsystem? Understanding these cross-subsystem resource conversion limits is essential for optimizing quantum algorithms in distributed architectures where local operations are restricted to real quantum channels.
	To address this fundamental question, we establish a rigorous framework for quantifying the distribution of coherence and imaginarity across bipartite systems. 
	
	\begin{definition}\label{Definition 2}
		Consider a bipartite Hilbert space $\mathcal{H}_A \otimes \mathcal{H}_B$ with fixed local computational bases. 
		The distance measures $\mathcal{D}$ employed in this distributed setting satisfy properties (i)--(iii) and additionally the condition of \emph{tensor consistency}: 
		$\mathcal{D}(\rho_X \otimes \tau_Y, \sigma_X \otimes \tau_Y) = \mathcal{D}(\rho_X, \sigma_X)$ for any auxiliary state $\tau_Y$, ensuring invariance under the attachment of uncorrelated ancillae.
		For a bipartite state $\rho_{AB} \in \mathfrak{D}(\mathcal{H}_A \otimes \mathcal{H}_B)$, the \emph{global coherence} is defined as the minimal distance to the set of fully incoherent bipartite states, 
		\begin{align}
			\mathcal{C}(\rho_{AB}) = \min_{\omega \in \mathcal{I}_{AB}} \mathcal{D}(\rho_{AB}, \omega). \nonumber
		\end{align}
		The \emph{$A$-local coherence} quantifies the coherence accessible in subsystem $A$ when $B$ is unrestricted, given by 
		\begin{align}
			\mathcal{C}_A(\rho_{AB}) = \min_{\delta \in \mathcal{I}_A} \mathcal{D}(\rho_{AB}, \delta),\nonumber
		\end{align}
		where $\mathcal{I}_A$ comprises states of the form $\sum_i p_i |i\rangle\langle i|_A \otimes \tau_B$ with arbitrary $\tau_B$. 
		The \emph{$B$-imaginarity} is evaluated solely on the reduced state $\rho_B = \mathrm{Tr}_A \rho_{AB}$ as 
		\begin{align}
			\mathcal{M}(\rho_B) = \min_{\sigma_B \in \mathcal{F}_B} \mathcal{D}(\rho_B, \sigma_B).\nonumber
		\end{align}
		Finally, let $\sigma_B^*$ and $\delta_B^*$ denote optimal states attaining the minima in the definitions of $\mathcal{M}(\rho_B)$ and the standard coherence measure $\mathcal{C}(\rho_B) = \min_{\delta \in \mathcal{I}_B} \mathcal{D}(\rho_B, \delta)$, respectively. 
		The \emph{residual $B$-coherence} is then 
		\begin{align}
			\mathcal{C}_R(\rho_B) = \mathcal{D}(\sigma_B^*, \delta_B^*),\nonumber
		\end{align}
		interpreted as the geometric separation between the closest real and the closest incoherent approximations of $\rho_B$. 
		If multiple minimizers exist, $\mathcal{C}_R(\rho_B)$ is taken as the infimum over all admissible pairs.
	\end{definition}
	
	\begin{theorem}\label{Theorem 2}
		For any bipartite quantum state $\rho_{AB}$ and metric $\mathcal{D}$ satisfying properties (i)--(iv), the following trade-off relation holds:
		\begin{align}
			\mathcal{C}_A(\rho_{AB})&\leq \mathcal{T}(\rho_{AB})+\Delta_A(\rho_A)+\mathcal{M}(\rho_B) + \mathcal{C}_R(\rho_B) \label{eq:bound1},
		\end{align}
		where $\rho_A=\mathrm{Tr}_B \rho_{AB}$, $\mathcal{T}(\rho_{AB})=\mathcal{D}(\rho_{AB},\rho_A\otimes \rho_B)$, and $\Delta_A(\rho_A)=\mathcal{D}(\rho_A,\rho_A^{\mathrm{diag}})$.
	\end{theorem}
	
	Theorem 2 establishes a fundamental allocation principle for quantum resources in distributed systems, see proof in Appendix A. The inequality reveals that the coherence achievable in subsystem $A$ is geometrically constrained by the following factors: $\Delta_A(\rho_A)$, which quantifies the intrinsic coherence of the reduced state of subsystem $A$; $\mathcal{T}(\rho_{AB})$, capturing the coherence enabled by total correlations between subsystems; and crucially, the internal structure of coherence within subsystem $B$, as characterized by the geometric framework introduced in Theorem 1. Specifically, the bound depends on $\mathcal{M}(\rho_B)$, which represents the convertible part of $B's$ coherence, and $\mathcal{C}_R(\rho_B)$, which represents the geometric conversion cost arising from the misalignment between $B's$ closest real and incoherent states. This decomposition underscores that it is not merely $B's$ total coherence $\mathcal{C}(\rho_B)$ that matters, but its composition, thereby delineating the fundamental limits of cross-subsystem resource utilization under real operations.
	
	To illustrate Theorem~2, we also employ the trace distance $\mathcal{D}_{\mathrm{tr}}(\alpha,\beta)=\|\alpha-\beta\|_{\mathrm{tr}}$ that satisfies (i)--(iv). Consider first a product state $\rho_{AB} = \rho_A \otimes \rho_B$, where
	\begin{align*}
		\rho_A = \begin{pmatrix}
			\frac{1}{2} & \frac{1}{4} \\[2pt]
			\frac{1}{4} & \frac{1}{2}
		\end{pmatrix}, \quad
		\rho_B = \begin{pmatrix}
			\frac{1}{2} & -\frac{i}{4} \\[2pt]
			\frac{i}{4} & \frac{1}{2}
		\end{pmatrix}.
	\end{align*}
	Direct computation yields
	$\mathcal{T}(\rho_{AB})=0$, $\Delta_A(\rho_A)=0.5$,
	$\mathcal{M}(\rho_B)=0.5$,
	$\mathcal{C}_R(\rho_B)=0$,
	$\mathcal{C}_A(\rho_{AB})=0.5$. 
	This confirms \eqref{eq:bound1}. The inequalities are strict in this case.
	
	For the entangled state $\rho_{AB} = p |\psi\rangle\langle\psi| + (1-p) \frac{\mathbb{I}_4}{4}$,
	where $|\psi\rangle = (|00\rangle + i |11\rangle)/\sqrt{2}$, $0 \leq p \leq 1$.
	Both reduced states $\rho_A$  and $\rho_B$ are proportional to identities. In this case, we have $\mathcal{T}(\rho_{AB})=\frac{3p}{2}$,
	$\Delta_A(\rho_A)=0$,
	$\mathcal{M}(\rho_B)=0$,
	$\mathcal{C}_R(\rho_B)=0$,
	$\mathcal{C}_A(\rho_{AB})=p$.
	The trade-off inequalitie \eqref{eq:bound1} exhibit strict inequality for all $0 < p \leq 1$, with equality attained only when $p=0$. These results illustrate the tightness of the derived bounds for entangled states, where non-zero correlations ($\mathcal{T} > 0$) impose fundamental limitations on the conversion from imaginarity into coherence.
	
	Theorem~2 reveals that, regardless of the specific form of the state, the coherence accessible in subsystem $A$ is fundamentally bounded by a combination of local coherence ($\Delta_A$), total correlations ($\mathcal{T}$), and—crucially—the internal geometric structure of subsystem $B$’s coherence, decomposed into its convertible imaginarity $\mathcal{M}(\rho_B)$ and the associated conversion cost $\mathcal{C}_R(\rho_B)$. This decomposition underscores that the limitation on cross‑subsystem resource conversion under real operations arises not merely from the amount of coherence present in $B$, but from its geometric composition. The strict inequality observed for the exemplary entangled state ($p>0$) illustrates an intrinsic activation barrier: even in the presence of non‑local correlations, full conversion of $B$’s imaginarity into $A$’s coherence is geometrically hindered. These insights provide a quantitative foundation for optimizing phase‑sensitive quantum protocols in distributed architectures, where the interplay of local resources, correlations, and geometric conversion costs dictates the ultimate efficiency of resource utilization.

	\section{Operational Protocols and Dynamical Limits}
\begin{figure}[t]
		\centering
		\includegraphics[width=\columnwidth]{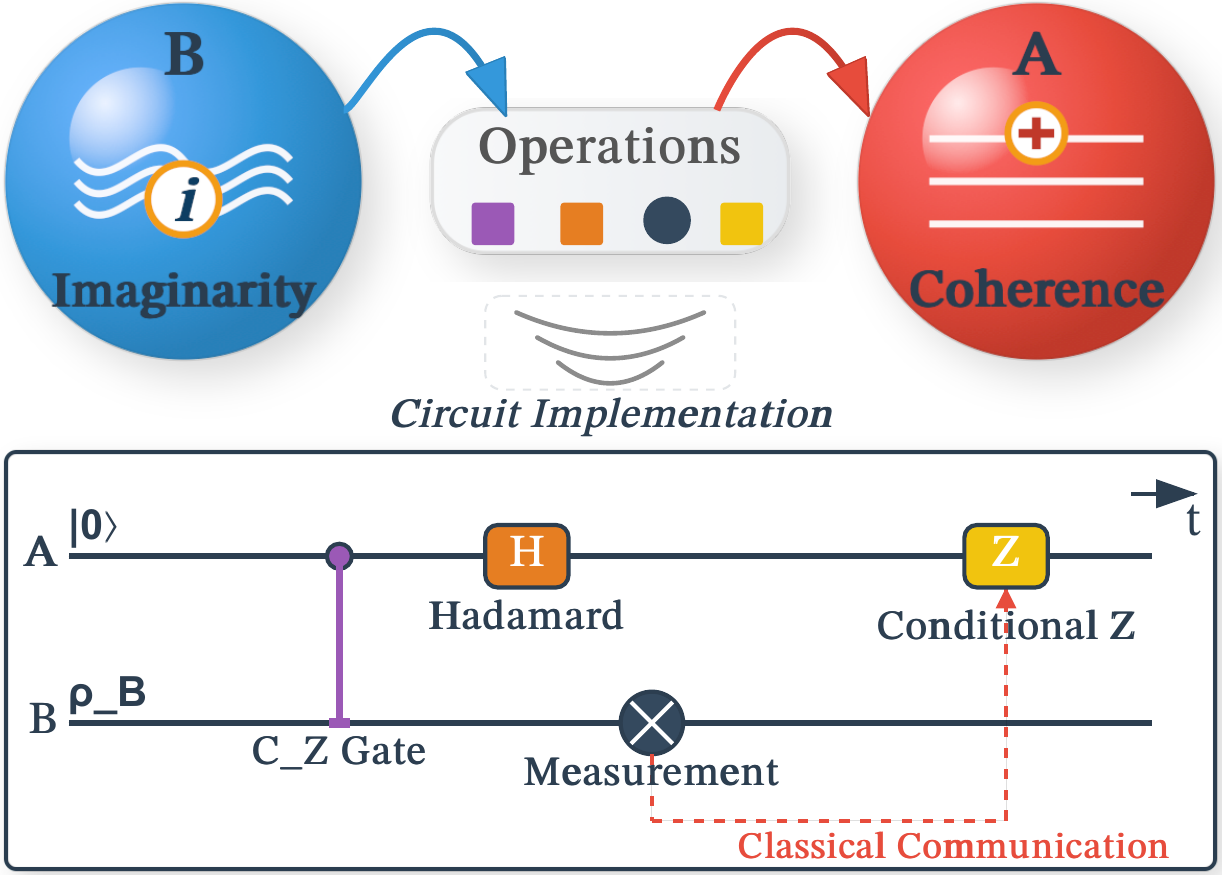}
		\caption{\textbf{Operational conversion of imaginarity into coherence.}  
			Upper panel: schematic illustration of the protocol converting imaginarity in subsystem $B$ into coherence in subsystem $A$ via real operations and a $Y$-basis measurement.  
			Lower panel: corresponding quantum circuit, consisting of a controlled-$Z$ gate, a Hadamard gate on $A$, a $Y$-basis measurement on $B$, and a conditional $Z_A$ correction (dashed red line).  
			The protocol achieves the exact equality $\mathcal{C}'_{\mathrm{tr}}(\rho_A^{\mathrm{final}}) = \mathcal{M}_{\mathrm{tr}}(\rho_B)$.}
		\label{IM3}
	\end{figure}

	A complete resource‑theoretic framework not only demands rigorous quantification of resources, but also requires the design of explicit protocols that can transform one resource into another under the free operations. Such protocols are crucial for demonstrating that the theoretical bounds derived earlier are physically attainable, and for enabling practical manipulation of quantum resources in real‑world settings.
	In distributed quantum systems, a key operational task is the conversion of imaginarity in one subsystem into coherence in another-a capability that is especially important when the available operations are restricted, for instance, to real quantum channels. Building on the geometric trade-off relations derived in Theorem 2, we construct a operational conversion protocol that transforms the imaginarity of subsystem $B$ into coherence of subsystem $A$ by combining real operations with a suitably chosen non‑real measurement, see Fig.~\ref{IM3}. This protocol provides an operational implementation of the abstract geometric relations, thereby testing the tightness of the theoretical bounds and laying a rigorous groundwork for investigating dynamical constraints and ultimate limits on resource conversion. The main result is stated in the following theorem; its proof is given in Appendix B.

	\begin{theorem}
		\label{Theorem 3}
		Consider a bipartite quantum system, where subsystem $A$ is initially prepared in the incoherent state $|0\rangle\langle 0|_A$, and subsystem $B$ is in an arbitrary single-qubit state $\rho_B$.
		There exists an operational protocol consisting of real operations, single-qubit unitaries, and a non-real projective measurement in the $Y$-basis, which converts the imaginarity of subsystem $B$ into coherence in subsystem $A$.
		The protocol proceeds as follows:
		\begin{enumerate}
			\item[(I)] Apply a controlled-$Z$ gate $C_Z = |0\rangle\langle 0|_A \otimes I_B + |1\rangle\langle 1|_A \otimes Z_B.$
			Since subsystem $A$ is initialized in $|0\rangle_A$, this operation acts trivially on the joint state: $C_Z (|0\rangle\langle 0|_A \otimes \rho_B) C_Z^\dagger
			= |0\rangle\langle 0|_A \otimes \rho_B.$
			Hence, no entanglement is generated at this stage.
			\item[(II)] Apply the Hadamard gate $H_A$ on subsystem $A$, producing $|+\rangle_A = \frac{1}{\sqrt{2}}(|0\rangle + |1\rangle).$
			\item[(III)] Measure subsystem $B$ in the $Y$-basis $\{|+i\rangle, |-i\rangle\}$ and communicate the outcome to Alice.
			\item[(IV)] Conditioned on the measurement outcome, Alice applies $Z_A$ if the result is $|-i\rangle$, and applies identity otherwise.
		\end{enumerate}
		Then this protocol exhibits the following properties:
		\begin{enumerate}
			\item[(a)] If $\rho_B$ is a real state ($\rho_B \in \mathcal{F}_B$), then the final state of subsystem $A$ is $\rho_A^{\mathrm{final}} = \mathbb{I}_A/2$, i.e., the maximally mixed state, which has zero coherence.
			\item[(b)] For any $\rho_B$ with non-zero imaginarity, the coherence of the final state satisfies
			\begin{align*}
				\mathcal{C}'_{\mathrm{tr}}(\rho_A^{\mathrm{final}}) = \mathcal{M}_{\mathrm{tr}}(\rho_B),
			\end{align*}
			thereby establishing an exact quantitative correspondence between the generated coherence and the initial imaginarity.
		\end{enumerate}
	\end{theorem}
	
	Theorem~3 provides an explicit operational conversion protocol that converts the imaginarity of subsystem~$B$ into coherence of subsystem~$A$. To further validate Theorem~3, we consider the following examples:
	
	We first consider the pure imaginary state $\rho_B = |+i\rangle\langle+i|_B$, which has maximal imaginarity $\mathcal{M}_{\mathrm{tr}}(\rho_B)=0.5$. With subsystem $A$ initialized in $|0\rangle_A$, the protocol yields outcome $|+i\rangle_B$ with unit probability, leaving $A$ in the coherent state $|+\rangle_A$. Consequently, $\mathcal{C}'_{\mathrm{tr}}(\rho_A^{\mathrm{final}})=0.5$, confirming the equality $\mathcal{C}'_{\mathrm{tr}}(\rho_A^{\mathrm{final}}) = \mathcal{M}_{\mathrm{tr}}(\rho_B)$. This demonstrates perfect conversion when the initial imaginarity is maximal.
	
	Next, consider the mixed single-qubit state
	\begin{align*}
		\rho_B = \begin{pmatrix} 0.7 & 0.2 + 0.3i \\ 0.2 - 0.3i & 0.3 \end{pmatrix},
	\end{align*}
	which has imaginarity $\mathcal{M}_{\mathrm{tr}}(\rho_B)=0.3$. Subsystem $A$ again starts in $|0\rangle_A$. Executing the protocol yields measurement probabilities $p_{+i}=0.2$, $p_{-i}=0.8$, and the final state of $A$ is
	\begin{align*}
		\rho_A^{\mathrm{final}} = 0.2|+\rangle\langle +|_A + 0.8|-\rangle\langle -|_A = \begin{pmatrix} 0.5 & -0.3 \\ -0.3 & 0.5 \end{pmatrix},
	\end{align*}
	resulting in $\mathcal{C}'_{\mathrm{tr}}(\rho_A^{\mathrm{final}})=0.3$. Thus, the equality $\mathcal{C}'_{\mathrm{tr}}(\rho_A^{\mathrm{final}}) = \mathcal{M}_{\mathrm{tr}}(\rho_B)$ holds exactly. This example illustrates that the protocol selectively converts the imaginary part of $B$'s coherence into an equivalent amount of trace distance coherence in $A$, while the real part of the off‑diagonal element does not contribute.
	
	Finally, to illustrate the necessity of imaginarity, we consider the case where $\rho_B$ is real. According to Theorem~3(a), for any real $\rho_B$, the protocol yields $\rho_A^{\mathrm{final}} = \mathbb{I}_A/2$. Specifically, consider an arbitrary single-qubit real state  
	\begin{align*}
		\rho_B = \begin{pmatrix} a & b \\ b & 1-a \end{pmatrix}, \quad a,b \in \mathbb{R}, \; 0 \leq a \leq 1.
	\end{align*}
	For such state, $\mathcal{M}_{\mathrm{tr}}(\rho_B)=0$. Under the protocol, the $Y$-basis measurement on $B$ yields outcome probabilities $p_{+i}=p_{-i}=\frac{1}{2}$, independent of $a$ and $b$. Consequently, the final state of $A$ is always maximally mixed, so $\mathcal{C}'_{\mathrm{tr}}(\rho_A^{\mathrm{final}})=0$. This confirms that coherence in $B$ alone is insufficient to generate coherence in $A$; only a nonzero imaginarity resource enables this conversion.
	
	Collectively, these examples confirm the precise correspondence established in Theorem~3: imaginarity in subsystem $B$ can be deterministically converted into an equivalent amount of coherence in subsystem $A$ through a combination of real operations and a specific non-real measurement. Theorem~3 thus provides an explicit operational protocol that realizes the geometric trade-offs outlined in Theorem~2, establishing the operational interconvertibility of imaginarity and coherence under physically admissible quantum operations.
	
	The operational significance of this conversion protocol becomes clear when one considers the distinct roles that imaginarity and coherence play in quantum information processing. While imaginarity provides unique advantages in tasks such as channel discrimination without ancillas \cite{KangDa} and Heisenberg limited multiparameter estimation \cite{Miyazaki}, coherence remains the canonical resource for a far broader class of quantum protocols. Notably, coherence has been established as the essential resource for quantum phase estimation, a core subroutine in algorithms including Shor's factoring algorithm and in quantum metrology \cite{Ahnefeld2025}. Recent work has demonstrated that every quantifiable unit of coherence directly improves phase estimation accuracy, establishing a direct operational link between a state's coherence and its utility in sensing and computational tasks \cite{Ahnefeld2025}. The conversion protocol thus enables imaginarity, which may be easier to generate under real operations in platforms with limited phase control \cite{Wu1}, to be transformed into coherence that can then be directly consumed in these widely applicable tasks. Moreover, since imaginarity can be obscured by the real components of a quantum state, the conversion serves a diagnostic function by making this hidden resource manifest as measurable coherence, facilitating resource verification in distributed quantum networks. The exact equality $\mathcal{C}'_{\mathrm{tr}}(\rho_A^{\mathrm{final}}) = \mathcal{M}_{\mathrm{tr}}(\rho_B)$ further demonstrates that the residual coherence term in Theorem~1 represents an irreducible conversion cost that cannot be eliminated under real operations, thereby giving operational meaning to the abstract geometric decomposition.
	
	While the operational conversion protocol in Theorem~3 demonstrates the interconversion of imaginarity and coherence under static operations, it is equally important to understand how these resources evolve dynamically. In closed quantum systems governed by a diagonal Hamiltonian, the conservation laws and oscillatory behaviors of coherence and imaginarity impose fundamental constraints on resource manipulation over time. Theorem~4 below provides a comprehensive analysis of these dynamical properties, revealing how the geometric trade-offs established in Section~II manifest under unitary evolution.
	
	\begin{theorem}\label{Theorem 4}
		For a single-qubit system evolving under a diagonal Hamiltonian $H = \frac{\omega}{2}\sigma_z$, the time-evolved state $\rho_t = e^{-iHt}\rho_0 e^{iHt}$ with $\rho_0 = \frac{1}{2}\bigl(I + x_0\sigma_x + y_0\sigma_y + z_0\sigma_z\bigr)$ exhibits exact dynamical conservation of modified trace-distance coherence, with complementary oscillations between imaginarity and residual coherence. These quantities satisfy the exact geometric constraint
		\begin{equation}
			\left(\frac{\mathcal{M}_{\mathrm{tr}}(\rho_t)}{\mathcal{C}'_{\mathrm{tr}}(\rho_0)}\right)^2 + \left(\frac{\mathcal{C}_{R_{\mathrm{tr}}}(\rho_t)}{\mathcal{C}'_{\mathrm{tr}}(\rho_0)}\right)^2 = 1,
		\end{equation}
		and the trade-off relation $\mathcal{C}'_{\mathrm{tr}}(\rho_t) \leq \mathcal{M}_{\mathrm{tr}}(\rho_t) + \mathcal{C}_{R_{\mathrm{tr}}}(\rho_t)$ holds for all $t$, with equality if and only if
		$\theta_0 - \omega t = n\frac{\pi}{2}$, where $n \in \mathbb{Z}$ and $\theta_0 = \arg(x_0 + i y_0)$.
	\end{theorem}
	
	Theorem~\ref{Theorem 4} establishes a comprehensive dynamical resource-theoretic framework that elucidates how coherence conservation and complementary oscillations between imaginarity and residual coherence govern quantum state evolution. The coherence conservation law underscores the inherent stability of quantum superpositions under diagonal Hamiltonian dynamics, while the phase-space constraints furnish a geometric interpretation of resource flow in the complex plane. The identified resonance conditions specify the fundamental prerequisites for temporal resource synchronization, thereby laying a rigorous theoretical foundation for optimizing time-dependent quantum protocols across varying system dimensions.
	
	Having characterized the intrinsic dynamical constraints on coherence and imaginarity, a natural subsequent question concerns the operational limits of resource conversion under quantum processes. Specifically, given that dynamical evolution preserves total coherence but permits oscillatory exchange between imaginarity and residual coherence, it is essential to determine how effectively imaginarity can be harnessed to generate coherence via physically admissible operations. This leads us to examine the maximum coherence $\mathcal{C}(\varepsilon(\rho))$ that can be extracted from a state $\rho$ with initial imaginarity $\mathcal{M}(\rho)$ under real operations $\varepsilon$. The following theorem quantifies this fundamental trade-off; the proof is provided in Appendix D.
	
	\begin{theorem}\label{Theorem 5}
		Let \(\rho \in \mathfrak{D}(\mathcal{H})\) be a quantum state and let \(\varepsilon: \mathfrak{D}(\mathcal{H}) \to \mathfrak{D}(\mathcal{H})\) be a real operation (i.e., \(\varepsilon(\mathcal{F}) \subseteq \mathcal{F}\)). Let \(\mathcal{D}\) be a contractive distance measure satisfying properties (i)--(iii) and property (v): for any completely positive trace‑preserving map \(\Phi\), \(\mathcal{D}(\Phi(\rho), \Phi(\sigma)) \leq \mathcal{D}(\rho, \sigma)\).  
		Let \(\sigma^* \in \mathcal{F}\) denote an optimal real state that achieves the minimum in the definition of \(\mathcal{M}(\rho) = \min_{\sigma \in \mathcal{F}} \mathcal{D}(\rho, \sigma)\).  
		Then, the coherence generated by the operation \(\varepsilon\) satisfies the inequality
		\begin{equation}\label{eq:conv_bound}
			\mathcal{C}(\varepsilon(\rho)) \leq \mathcal{M}(\rho) + \mathcal{C}(\varepsilon(\sigma^*)).
		\end{equation}
	\end{theorem}
	
	Theorem~5 establishes a general upper bound on the coherence that can be generated from a state with a given amount of imaginarity under real operations. This bound consists of two parts: the initial imaginarity $\mathcal{M}(\rho)$ and the coherence that can be generated from the closest real state $\mathcal{C}(\varepsilon(\sigma^*))$. The first part reflects the convertible portion of the resource, while the second part represents the coherence that the operation $\varepsilon$ can produce even from a real state, which is independent of the imaginarity of $\rho$.
	The bound in Theorem~5 is universal for any contractive distance measure, including the trace distance and the Bures distance, which are widely used in resource theories. It provides a fundamental limit on the efficiency of converting imaginarity into coherence under real operations. In particular, if an operation $\varepsilon$ cannot generate coherence from real states (i.e., $\mathcal{C}(\varepsilon(\sigma^*))=0$ for all real $\sigma^*$), then the generated coherence is at most the initial imaginarity. This is the case for strictly incoherent operations, which map incoherent states to incoherent states. On the other hand, if the operation can generate coherence from real states, the bound allows for more coherence than the initial imaginarity, but this extra coherence is attributed to the operation itself rather than the initial resource.
	Theorem~5 complements the geometric decomposition established in Theorem~1 in the following sense. Theorem~1 identifies a purely state‑dependent constraint: the total coherence of a fixed state is bounded by the sum of its imaginarity and a residual geometric term $\mathcal{C}_R$. Theorem~5, in contrast, characterizes a channel‑dependent constraint: under a real operation $\varepsilon$, the coherence of the output state is bounded by the initial imaginarity plus the coherence that $\varepsilon$ itself can generate from a real input. Thus, the two results address distinct facets of the coherence–imaginarity relationship---one static and geometric, the other dynamic and operational. Together they furnish a complete picture of the limitations on resource interconversion.
	The operational protocol in Theorem~3 achieves the equality $\mathcal{C}'_{\mathrm{tr}}(\rho_A^{\mathrm{final}}) = \mathcal{M}_{\mathrm{tr}}(\rho_B)$, corresponding to the case $\mathcal{C}(\varepsilon(\sigma^*))=0$ in Theorem~5. The dynamical evolution governed by a diagonal Hamiltonian, as analyzed in Theorem~4, also satisfies the bound in Theorem~5 with equality, since unitary evolution under a real Hamiltonian does not generate coherence from real states. These consistencies underscore the coherence of the overall framework.
	So, Theorem~5 provides a general and powerful tool for analyzing the interconversion between imaginarity and coherence under real operations, setting fundamental limits for resource conversion in quantum information processing.
	
	\section{Discussion}
	The concept of residual coherence introduced in this work provides a geometric pivot for disentangling the intertwined relationship between coherence and imaginarity. In contrast to earlier studies that established the formal resource theory of imaginarity or developed specific quantification methods, our framework offers a structural decomposition of coherence itself. Instead of treating imaginarity as an independent resource merely coexisting with coherence, we identify it as a separable constituent within the coherence hierarchy. The central inequality reveals that total coherence comprises a purely imaginary contribution and a residual term originating from the misalignment between the state's real components and the preferred basis. This geometric picture explains why two states with identical total coherence may exhibit markedly different conversion capabilities under restricted operations: the distinction resides in their internal partition between imaginarity and residual coherence.
	The geometric decomposition carries immediate operational and dynamical consequences. First, it sets fundamental conversion limits. Our explicit protocol achieves exact conversion only when the residual term vanishes, whereas the general bound establishes the residual part as the irreducible geometric overhead for arbitrary states. This shifts the focus from monotonicity of transformations to a direct link between conversion efficiency and the intrinsic geometry of the state space. Second, under diagonal Hamiltonian evolution, we find that total coherence is strictly conserved while imaginarity and residual coherence undergo complementary oscillations governed by a circular constraint. The channel-dependent bound derived in Theorem~5 complements these state-centric insights: under any real operation, the extractable coherence is limited by the sum of the initial imaginarity and the coherence that the operation itself can generate from a real state. Together, these results unify static geometric constraints with dynamical operational limits and furnish a complete picture of coherence-imaginarity interconversion.
	
	From a practical standpoint, the geometric framework acquires concrete meaning in linear optical architectures, where generating complex amplitudes demands additional wave plates relative to real operations. The imaginarity resource also underpins the advantage in specific distributed tasks, including ancilla-free channel discrimination and Heisenberg-limited multiparameter estimation. The conversion protocol presented here is directly implementable on existing photonic or trapped-ion platforms, as it employs only a controlled-\(Z\) gate, single-qubit rotations, and a \(Y\)-basis measurement. By transforming imaginarity, which may be easier to prepare or store under phase-restricted conditions, into coherence, the protocol activates a resource that can be consumed in a broader class of applications, most notably quantum phase estimation. The exact conversion equality further demonstrates that residual coherence acts as a geometric diagnostic: any shortfall in generated coherence signals a nonzero residual term and thus an intrinsic reference-frame mismatch.
	
	In summary, we have established a unified geometric framework that clarifies the hierarchical structure of quantum coherence, quantifies the cross-subsystem conversion limits in bipartite settings, and provides an explicit, experimentally viable protocol for converting imaginarity into usable coherence. This work not only elucidates the constitutive role of imaginarity but also supplies quantitative bounds for resource management in distributed quantum technologies.
	Looking ahead, several directions merit further investigation. Extending the geometric decomposition to multipartite systems could uncover how collective network phenomena depend on the distribution of imaginarity and residual coherence. An analysis in continuous-variable settings may lead to optimized manipulation protocols for Gaussian resources. The derived conversion bounds also offer a theoretical foundation for resource allocation in distributed quantum computing and metrology, while the geometric characterization of coherence components could inspire new strategies for quantum error correction. As quantum technologies progress toward scalable architectures, the fundamental limits and geometric principles developed here should serve as essential navigational tools for resource optimization across diverse physical platforms.

	\bigskip
	
	\noindent{\bf Acknowledgments}\, \,
	We sincerely thank Professor Xu Jianwei for his profound and appropriate comments on revising our results,
	This work was supported by the National Natural Science Foundation of China under Grant 12175147; the Jiangxi Provincial Natural Science Foundation under Grant 20252BAC200156; the Zhejiang Provincial Natural Science Foundation under Grant LZ24A050005; the Early-Career Young Scientists and Technologists Project of Jiangxi Province under Grant 20244BCE52197; the Science and Technology Project of the Jiangxi Provincial Department of Education under Grants GJJ2400606, GJJ2200730; the Doctoral Research Startup Fund Project of East China University of Technology under Grant DHBK2024029; and the specific research fund of the Innovation Platform for Academicians of Hainan Province under Grant YSPTZX202215.

	\section*{data availability}
	There are no publicly available research data or software supporting this paper. Requests for further information or data should be sent to the authors.

	\appendix
	\section{Proof of Theorem 2}
	Consider the candidate state $\sigma^* = \rho_A^{\mathrm{diag}} \otimes \delta_B^* \in \mathcal{I}_A$.
	By the definition of the $A$-local coherence, we have
	\begin{equation}
		\mathcal{C}_A(\rho_{AB}) \leq \mathcal{D}(\rho_{AB}, \rho_A^{\mathrm{diag}} \otimes \delta_B^*). \label{eq:core}
	\end{equation}
	Applying the triangle inequality with the intermediate state $\rho_A \otimes \rho_B$ and $\rho_A^{\mathrm{diag}} \otimes \rho_B$, we obtain
	\begin{align}
		\mathcal{D}(\rho_{AB}, \rho_A^{\mathrm{diag}} \otimes \delta_B^*)\leq& \mathcal{D}(\rho_{AB}, \rho_A \otimes \rho_B) \label{eq:tri1}\\
		&+ \mathcal{D}(\rho_A \otimes \rho_B, \rho_A^{\mathrm{diag}} \otimes \delta_B^*),\nonumber\\
		\mathcal{D}(\rho_A \otimes \rho_B, \rho_A^{\mathrm{diag}} \otimes \delta_B^*) \leq& \mathcal{D}(\rho_A \otimes \rho_B, \rho_A^{\mathrm{diag}} \otimes \rho_B)\label{eq:tri2} \\
		&+ \mathcal{D}(\rho_A^{\mathrm{diag}} \otimes \rho_B, \rho_A^{\mathrm{diag}} \otimes \delta_B^*). \nonumber
	\end{align}
	It follows from the tensor consistency property (iv) that
	\begin{align*}
		\mathcal{D}(\rho_A \otimes \rho_B, \rho_A^{\mathrm{diag}} \otimes \rho_B) &= \mathcal{D}(\rho_A, \rho_A^{\mathrm{diag}}) = \Delta_A(\rho_A), \\
		\mathcal{D}(\rho_A^{\mathrm{diag}} \otimes \rho_B, \rho_A^{\mathrm{diag}} \otimes \delta_B^*) &= \mathcal{D}(\rho_B, \delta_B^*) = \mathcal{C}(\rho_B).
	\end{align*}
	By construction, $\delta_B^*$ is precisely the optimal incoherent state $\rho_d$ for subsystem $B$ in the decomposition of Theorem~1. Applying that theorem to $\rho_B$ yields $\mathcal{C}(\rho_B) \leq \mathcal{M}(\rho_B)+\mathcal{C}_R(\rho_B)$. Combining inequalities \eqref{eq:core} -- \eqref{eq:tri2}, we arrive at
	\begin{align*}
		\mathcal{C}_A(\rho_{AB}) 
		&\leq \mathcal{T}(\rho_{AB}) + \Delta_A(\rho_A) + \mathcal{C}(\rho_B) \\
		&\leq \mathcal{T}(\rho_{AB}) + \Delta_A(\rho_A) + \mathcal{M}(\rho_B) + \mathcal{C}_R(\rho_B).
	\end{align*}
	This completes the proof.
	
	\section{Proof of Theorem 3}
	We begin by expressing the initial state of subsystem $B$ in the computational basis:
	\begin{align*}
		\rho_B = \begin{pmatrix} p & \alpha \\ \alpha^* & 1-p \end{pmatrix}, \qquad 
		\alpha = x + i y, \quad x, y \in \mathbb{R}, \quad 0 \le p \le 1.
	\end{align*}
	Subsystem $A$ is prepared in the incoherent state $|0\rangle\langle 0|_A$. The protocol proceeds as follows:
	
	Step I: Apply the controlled-$Z$ gate $C_Z = |0\rangle\langle 0|_A \otimes I_B + |1\rangle\langle 1|_A \otimes Z_B$. Since the control qubit $A$ is initially in $|0\rangle$, the joint state remains unchanged:
	\begin{align*}
		\rho^{(1)}_{AB}=C_Z\bigl(|0\rangle\langle 0|_A\otimes\rho_B\bigr)C_Z^\dagger = |0\rangle\langle 0|_A\otimes\rho_B .
	\end{align*}
	
	Step II: Apply the Hadamard gate $H_A$ to qubit $A$. Using $H|0\rangle = |+\rangle$ where $|+\rangle = (|0\rangle+|1\rangle)/\sqrt{2}$, we obtain
	\begin{align*}
		\rho^{(2)}_{AB}= (H_A\otimes I_B)\rho^{(1)}_{AB}(H_A\otimes I_B)^\dagger = |+\rangle\langle +|_A\otimes\rho_B .
	\end{align*}
	
	Step III: Bob measures subsystem $B$ in the $Y$-basis $\{|{+i}\rangle_B,|{-i}\rangle_B\}$, where $|{\pm i}\rangle_B = (|0\rangle_B \pm i|1\rangle_B)/\sqrt{2}$. The corresponding measurement operators are $M_{+i}=|{+i}\rangle\langle{+i}|_B$ and $M_{-i}=|{-i}\rangle\langle{-i}|_B$. The probability for outcome $\pm i$ is
	\begin{align*}
		p_{\pm i}= \operatorname{Tr}\bigl[(I_A\otimes M_{\pm i})\,\rho^{(2)}_{AB}\bigr]
		= \langle{\pm i}|\rho_B|{\pm i}\rangle .
	\end{align*}
	A direct computation gives
	\begin{align*}
		\langle{+i}|\rho_B|{+i}\rangle &= \frac12\Bigl(1 + i(\alpha-\alpha^*)\Bigr) = \frac12(1-2y), \\
		\langle{-i}|\rho_B|{-i}\rangle &= \frac12\Bigl(1 - i(\alpha-\alpha^*)\Bigr) = \frac12(1+2y),
	\end{align*}
	so that $p_{+i}= \frac12 - y$ and $p_{-i}= \frac12 + y$. The positivity of $\rho_B$ implies $|\alpha|^2 \leq p(1-p) \leq \frac14$, hence $|y| \leq \frac12$, ensuring both probabilities are non-negative.
	After the measurement (before the conditional correction), the unnormalized state of $A$ corresponding to outcome $\pm i$ is $\widetilde{\rho}^{(\pm i)}_A = p_{\pm i}\,|+\rangle\langle +|_A$.
	
	Step IV: Conditioned on the outcome: if the result is ${-i}$, Alice applies the Pauli-$Z$ gate to her qubit; otherwise she does nothing. Consequently, after the correction, the normalized state of $A$ for outcome ${+i}$ is $|+\rangle\langle +|_A$, and for outcome ${-i}$ it is $Z|+\rangle\langle +|Z = |-\rangle\langle -|_A$. The average final state of subsystem $A$ is therefore
	\begin{align*}
		\rho_A^{\mathrm{final}} = p_{+i}\,|+\rangle\langle +|_A \;+\; p_{-i}\,|-\rangle\langle -|_A .
	\end{align*}
	Using $|+\rangle\langle +| = \frac12(I+X)$ and $|-\rangle\langle -| = \frac12(I-X)$, where $I$ is the identity and $X$ the Pauli-$x$ matrix, we obtain
	\begin{align}\label{eq:finalState}
		\rho_A^{\mathrm{final}} &= \frac12\Bigl[\,I + (p_{+i}-p_{-i})X\,\Bigr]\nonumber\\
		&= \frac12\bigl(I - 2y\,X\bigr)\nonumber\\
		&= \frac12\begin{pmatrix}1 & -2y \\ -2y & 1\end{pmatrix}.
	\end{align}
	Thus the final state depends solely on the imaginary part $y$ of the off‑diagonal element $\alpha$; the real part $x$ and the population $p$ do not appear.
	
	We now evaluate the modified trace‑distance coherence of $\rho_A^{\mathrm{final}}$. By definition,
	\begin{align*}
		\mathcal{C}'_{\mathrm{tr}}(\rho) = \min_{\lambda\ge 0,\; \delta\in\mathcal{I}} \|\rho-\lambda\delta\|_{\mathrm{tr}},
	\end{align*}
	where $\mathcal{I}$ denotes the set of incoherent states. For a qubit, any incoherent state can be written as $\delta = \frac12(I + s_z Z)$ with $s_z\in[-1,1]$. With $\lambda\ge 0$, the difference matrix is
	\begin{align*}
		\rho_A^{\mathrm{final}} - \lambda\delta
		= \frac12\bigl[ (1-\lambda)I - 2yX - \lambda s_z Z \bigr].
	\end{align*}
	Its eigenvalues are $\frac12\bigl[(1-\lambda) \pm \sqrt{4y^{2} + \lambda^{2}s_z^{2}}\bigr]$, so
	\begin{align*}
		\bigl\|\rho_A^{\mathrm{final}} - \lambda\delta\bigr\|_{\mathrm{tr}}
		= \max\Bigl\{ |1-\lambda|,\; \sqrt{4y^{2} + \lambda^{2}s_z^{2}} \Bigr\}.
	\end{align*}
	For fixed $\lambda$, the square‑root term is minimized by choosing $s_z=0$. Hence the joint minimization reduces to
	\begin{align*}
		\mathcal{C}'_{\mathrm{tr}}(\rho_A^{\mathrm{final}})
		= \min_{\lambda\ge 0} \max\bigl\{ |1-\lambda|,\; 2|y| \bigr\}.
	\end{align*}
	The function $f(\lambda)=\max\{|1-\lambda|,\,2|y|\}$ attains its minimum when $|1-\lambda| = 2|y|$, achieved by taking $\lambda^* = 1-2|y|$ (which is non‑negative because $|y|\le\frac12$). At this optimum, $f(\lambda^*) = 2|y|$, yielding
	\begin{align*}
		\mathcal{C}'_{\mathrm{tr}}(\rho_A^{\mathrm{final}}) = 2|y|.
	\end{align*}
	
	Next, we compute the trace‑distance imaginarity of the initial state $\rho_B$. For a single‑qubit state, a closest real state $\sigma^*$ is obtained by keeping the diagonal elements and replacing the off‑diagonal entry with its real part:
	\begin{align*}
		\sigma^* = \begin{pmatrix} p & x \\ x & 1-p \end{pmatrix}, \qquad x = \Re(\alpha).
	\end{align*}
	Then $\rho_B - \sigma^* = \begin{pmatrix} 0 & iy \\ -iy & 0 \end{pmatrix}$, whose non‑zero eigenvalues are $\pm i|y|$, giving $\|\rho_B - \sigma^*\|_{\mathrm{tr}} = 2|y|$. By definition, $\mathcal{M}_{\mathrm{tr}}(\rho_B) = \min_{\sigma\in\mathcal{F}}\|\rho_B-\sigma\|_{\mathrm{tr}} \le 2|y|$. Moreover, for qubits the equality $\mathcal{M}_{\mathrm{tr}}(\rho_B)=\frac12\|\rho_B-\rho_B^{T}\|_{\mathrm{tr}}$ holds \cite{Hickey1}. Since
	\begin{align*}
		\rho_B - \rho_B^{T} = \begin{pmatrix} 0 & 2iy \\ -2iy & 0 \end{pmatrix},
	\end{align*}
	we have $\|\rho_B-\rho_B^{T}\|_{\mathrm{tr}} = 4|y|$. Consequently,
	\begin{align*}
		\mathcal{M}_{\mathrm{tr}}(\rho_B) = 2|y|.
	\end{align*}
	
	Combining the two results, we obtain the exact relation
	\begin{align*}
		\mathcal{C}'_{\mathrm{tr}}(\rho_A^{\mathrm{final}}) = 2|y| = \mathcal{M}_{\mathrm{tr}}(\rho_B),
	\end{align*}
	which establishes part (b) of the theorem.
	
	Finally, if $\rho_B$ is real, then $y=0$. From \eqref{eq:finalState} we have $\rho_A^{\mathrm{final}} = \frac12 I$, the maximally mixed state. Evaluating the modified trace‑distance coherence for this state, we may take $\lambda=1$ and $\delta = \frac12 I$ (which is incoherent), yielding $\|\frac12 I - 1\cdot\frac12 I\|_{\mathrm{tr}} = 0$. Hence $\mathcal{C}'_{\mathrm{tr}}(\rho_A^{\mathrm{final}}) = 0$, confirming part (a) of the theorem: no coherence is generated when the initial imaginarity vanishes.

	\section{Proof of Theorem 4}
	The initial state in the Bloch representation is $\rho_0 = \frac{1}{2}(I + x_0\sigma_x + y_0\sigma_y + z_0\sigma_z)$. Under the Hamiltonian $H = \frac{\omega}{2}\sigma_z$, the evolution operator $U_t = e^{-iHt} = \text{diag}(e^{-i\omega t/2}, e^{i\omega t/2})$ generates the state
	\begin{align*}
		\rho_t &= U_t \rho_0 U_t^\dagger\\
		& = \frac{1}{2}\begin{pmatrix}
			1+z_0 & (x_0 - i y_0)e^{-i\omega t} \\
			(x_0 + i y_0)e^{i\omega t} & 1-z_0
		\end{pmatrix}.
	\end{align*}
	The off-diagonal element can be expressed as $(\rho_t)_{01} = \frac{r_\perp}{2} e^{-i(\omega t - \theta_0)}$, where $r_\perp = \sqrt{x_0^2 + y_0^2}$ and $\theta_0 = \arg(x_0 + i y_0)$.
	
	To compute the modified trace-distance coherence $\mathcal{C}'_{\mathrm{tr}}(\rho_t)$, we consider the optimization over scaling factors $\lambda \geq 0$ and incoherent states $\delta = \text{diag}(p, 1-p)$. Direct calculation of the trace norm $\|\rho_t - \lambda\delta\|_{\mathrm{tr}}$ yields
	\begin{align*}
		\|\rho_t - \lambda\delta\|_{\mathrm{tr}}
		=
		\max\left\{
		|1-\lambda|,
		\sqrt{r_\perp^2+\bigl(z_0-\lambda(2p-1)\bigr)^2}
		\right\}.
	\end{align*}
	Hence $\|\rho_t - \lambda\delta\|_{\mathrm{tr}}\ge r_\perp.$
	The lower bound is attained by choosing
	\(\lambda=1\) and \(p=(1+z_0)/2\), so that
	\(\delta=\rho_t^{\mathrm{diag}}\). Therefore,
	$\mathcal{C}'_{\mathrm{tr}}(\rho_t) = r_\perp = \mathcal{C}'_{\mathrm{tr}}(\rho_0)$ for all $t$.
	
	For the trace-distance imaginarity $\mathcal{M}_{\mathrm{tr}}(\rho_t)$, we minimize over real states $\sigma \in \mathcal{F}$. In the Bloch representation, real states correspond to vectors of the form $(x,0,z)$. The closest real state to $\rho_t$ is obtained by setting the $y$-component to zero, yielding $\sigma_t = \frac{1}{2}(I + x_t\sigma_x + z_0\sigma_z)$ where $x_t = r_\perp\cos(\theta_0 - \omega t)$. The trace distance between $\rho_t$ and $\sigma_t$ is
	\begin{align*}
		\|\rho_t - \sigma_t\|_{\mathrm{tr}} = |y_t| = r_\perp |\sin(\theta_0 - \omega t)|,
	\end{align*}
	where $y_t = r_\perp\sin(\theta_0 - \omega t)$ is the $y$-component of $\rho_t$'s Bloch vector. Therefore, $\mathcal{M}_{\mathrm{tr}}(\rho_t) = r_\perp |\sin(\theta_0 - \omega t)|$.
	
	The residual coherence $\mathcal{C}_{R_{\mathrm{tr}}}(\rho_t)$ is defined as the trace distance between the optimal real state $\sigma_t$ (from the imaginarity minimization) and the optimal scaled incoherent state $\rho_t^{\mathrm{diag}}$ (from the coherence minimization). Since $\sigma_t = \frac{1}{2}(I + x_t\sigma_x + z_0\sigma_z)$ and $\rho_t^{\mathrm{diag}} = \frac{1}{2}(I + z_0\sigma_z)$, their difference is
	\begin{align*}
		\sigma_t - \rho_t^{\mathrm{diag}} = \frac{x_t}{2}\sigma_x,
	\end{align*}
	whose trace norm is $|x_t| = r_\perp |\cos(\theta_0 - \omega t)|$. Thus, $\mathcal{C}_{R_{\mathrm{tr}}}(\rho_t) = r_\perp |\cos(\theta_0 - \omega t)|$.
	
	The expressions for $\mathcal{M}_{\mathrm{tr}}(\rho_t)$ and $\mathcal{C}_{R_{\mathrm{tr}}}(\rho_t)$ immediately yield the geometric constraint
	\begin{align*}
		&\left(\frac{\mathcal{M}_{\mathrm{tr}}(\rho_t)}{r_\perp}\right)^2 + \left(\frac{\mathcal{C}_{R_{\mathrm{tr}}}(\rho_t)}{r_\perp}\right)^2\\
		& = \sin^2(\theta_0 - \omega t) + \cos^2(\theta_0 - \omega t) = 1.
	\end{align*}
	
	Finally, the trade-off inequality $\mathcal{C}'_{\mathrm{tr}}(\rho_t) \leq \mathcal{M}_{\mathrm{tr}}(\rho_t) + \mathcal{C}_{R_{\mathrm{tr}}}(\rho_t)$ becomes
	\begin{align*}
		r_\perp \leq r_\perp\bigl(|\sin(\theta_0 - \omega t)| + |\cos(\theta_0 - \omega t)|\bigr).
	\end{align*}
	Since $|\sin\phi| + |\cos\phi| \geq 1$ for all real $\phi$, equality holds if and only if $|\sin\phi| + |\cos\phi| = 1$, which is equivalent to $\sin\phi \cos\phi = 0$, i.e., $\phi = n\pi/2$ for some integer $n$. Setting $\phi = \theta_0 - \omega t$ completes the proof.

	\section{Proof of Theorem 5}
	By the contractivity of the distance measure $\mathcal{D}$ (Property (v)), we have
	\begin{equation}\label{eq:contract}
		\mathcal{D}(\varepsilon(\rho), \varepsilon(\sigma^{*})) \leq \mathcal{D}(\rho, \sigma^{*}) = \mathcal{M}(\rho).
	\end{equation}
	For any incoherent state $\delta \in \mathcal{I}$, the triangle inequality (Property (iii)) implies
	$\mathcal{D}(\varepsilon(\rho), \delta) \leq \mathcal{D}(\varepsilon(\rho), \varepsilon(\sigma^{*})) + \mathcal{D}(\varepsilon(\sigma^{*}), \delta)$. Therefore,
	\begin{align*}
		\mathcal{C}(\varepsilon(\rho)) &= \min_{\delta \in \mathcal{I}} \mathcal{D}(\varepsilon(\rho), \delta) \\
		&\leq \mathcal{D}(\varepsilon(\rho), \varepsilon(\sigma^{*})) + \min_{\delta \in \mathcal{I}} \mathcal{D}(\varepsilon(\sigma^{*}), \delta) \\
		&= \mathcal{D}(\varepsilon(\rho), \varepsilon(\sigma^{*})) + \mathcal{C}(\varepsilon(\sigma^{*})) \\
		&\leq \mathcal{M}(\rho) + \mathcal{C}(\varepsilon(\sigma^{*})),
	\end{align*}
	where the final inequality follows directly from \eqref{eq:contract}. This completes the proof.

\end{document}